\documentclass[11pt]{article}
\usepackage{amsmath}
\usepackage{latexsym}
\usepackage{amsfonts}
\usepackage{amssymb}
\newtheorem{lemma}{Lemma}
\newtheorem{definition}{Definition}

\newtheorem{proposition}{Proposition}

\newcommand{\al}{\alpha}
\newcommand{\be}{\beta}
\newcommand{\ga}{\gamma}
\newcommand{\de}{\delta}
\newcommand{\ep}{\epsilon}
\newcommand{\om}{\omega}
\newcommand{\+}{\!+\!}
\newcommand{\m}{\!-\!}
\newcommand{\half}{{\lower0.25ex\hbox{\raise.6ex\hbox{\the\scriptfont0 1}\kern-.2em\slash\kern-.1em\lower.25ex\hbox{\the\scriptfont0 2}}}}
\newcommand{\thalf}{{\lower0.25ex\hbox{\raise.6ex\hbox{\the\scriptfont0 3}\kern-.2em\slash\kern-.1em\lower.25ex\hbox{\the\scriptfont0 2}}}}
\newcommand{\fhalf}{{\lower0.25ex\hbox{\raise.6ex\hbox{\the\scriptfont0 5}\kern-.2em\slash\kern-.1em\lower.25ex\hbox{\the\scriptfont0 2}}}}
\newcommand{\quar}{{\lower0.25ex\hbox{\raise.6ex\hbox{\the\scriptfont0 1}\kern-.2em/\kern-.1em\lower.25ex\hbox{\the\scriptfont0 4}}}}

\renewcommand{\arraystretch}{1.5}
\begin{document}
\title{
\large\bf
Gap probabilities in the finite and scaled Cauchy random matrix ensembles}
\author
{\large N.S. Witte,\\
{\it Department of Mathematics and Statistics}\\
{\&} {\it School of Physics, University of Melbourne}\\ 
{\it Parkville, Victoria 3052, AUSTRALIA}\\
and {\large P.J. Forrester}\\
{\it Department of Mathematics and Statistics}\\
{\it University of Melbourne}\\
{\it Parkville, Victoria 3052, AUSTRALIA.} }
\maketitle
\begin{abstract}
The probabilities for gaps in the eigenvalue spectrum of finite $ N\times N $
random unitary ensembles on the unit circle with a singular weight, and 
the related hermitian ensembles on the line with Cauchy weight, are found
exactly. The finite cases for exclusion from single and double intervals are
given in terms of second order second degree ODEs which are related to
certain \mbox{Painlev\'e-VI} transcendents. The scaled cases in the
thermodynamic limit are again second degree and second order, this time related
to \mbox{Painlev\'e-V} transcendents. Using transformations relating
the second degree ODE and transcendent we prove an identity for the
scaled bulk limit which leads to a simple expression for the spacing p.d.f.
We also relate all the variables appearing in the Fredholm determinant formalism
to particular \mbox{Painlev\'e} transcendents, in a simple and transparent
way, and exhibit their scaling behaviour.
\end{abstract}
\vfill\eject
\section{Introduction}

Our intention in this work is to study two intimately related ensembles of
random matrices with unitary symmetry - the circular Jacobi unitary ensemble
consisting of eigenvalues confined to the unit circle with a spectrum singularity,
and the Cauchy ensemble on the real line. 
As we shall see, there is a mapping between these two ensembles which allows 
us to find all the results for the circular Jacobi ensemble from those of the 
Cauchy ensemble. The latter is a classical matrix ensemble on the real line,
and as such there is a well established method for characterising the gap 
probabilities in terms of differential equations \cite{TW-94}. This avoids the 
clumsy and unnecessary work of recasting all of the formalism onto the unit 
circle. The circular Jacobi ensemble that we consider is one with an 
algebraic singularity in its spectrum, which in the log-gas picture of the
eigenvalue p.d.f. can be interpreted as an impurity with variable charge
fixed at $ z = 1 $ interacting with the mobile unit charges representing the
eigenvalues.

In general an ensemble of unitary matrices with unitary symmetry has an 
eigenvalue p.d.f. of the form
\begin{equation}
  \prod_{l=1}^N w_2(z_l) \prod_{1 \le j < k \le N} | z_k - z_j |^2 \ ,
\label{cJUE-pdf}
\end{equation}
$ z = e^{i\theta} = e^{2\pi ix/L}, 
  \quad \theta \in [0,2\pi), \quad x \in [0,L) $, while an ensemble of 
Hermitian matrices with unitary symmetry has an eigenvalue p.d.f. of the form
\begin{equation}
  \prod_{l=1}^N w_2(\lambda_l) \prod_{1 \le j < k \le N}
  | \lambda_k - \lambda_j |^2 \ .
\label{CyUE-pdf}
\end{equation}
The circular Jacobi ensemble with unitary symmetry, denoted as cJUE, is
specified by the p.d.f. (\ref{cJUE-pdf}) with the weight function
\begin{equation}
  w_2(z) = |1-z|^{2a} \ .
\label{cJUE-weight}
\end{equation}
In the case $ a = 1 $ this is realised by the eigenvalues $ z = e^{i\theta} $
from the ensemble of $ (N+1)\times(N+1) $ random unitary matrices, with
all angles $ \theta $ measured from any one eigenvalue. The Cauchy unitary 
ensemble, denoted by CyUE, is specified by (\ref{CyUE-pdf}) with
\begin{equation}
  w_2(\lambda) = (1\+\lambda^2)^{-N-a}, \quad \lambda \in (-\infty,\infty) \ .
\label{CyUE-weight}
\end{equation}
It results from the cJUE by mapping the eigenvalues from the unit circle to
the line tangent to its south-most point via a stereographic projection from
its north-most point. Specifying points on the unit circle via an azimuthal
angle $ \theta $, so that $ \theta = 0 $ corresponds to the north-most point
and $ \theta = -\pi,\pi $ to the south-most point, this is achieved by the
mapping
\begin{equation}
   \lambda = \cot(\theta/2) \ ,
\end{equation}
and we find
\begin{multline}
     \prod^N_{i=1} |1\m z_i|^{2a} \prod_{1 \leq j < k \leq N}|z_k\m z_j|^2
      d\theta_1 \ldots d\theta_N \\
  \propto \, \prod^N_{i=1} {1 \over (1\+ \lambda^2_i)^{N+a}}
             \prod_{1 \leq j < k \leq N} |\lambda_k\m \lambda_j|^2 
             d\lambda_1 \ldots d\lambda_N \ .
\label{eigen-pdf-xfm}
\end{multline}

In (\ref{CyUE-pdf}), let us write
\begin{equation}
   w_2(\lambda) = e^{-2V(\lambda)} \quad\text{and}\quad 
   2V'(\lambda) = {g(\lambda) \over f(\lambda)} \ ,
\label{exp-weight}
\end{equation}
with $ f(\lambda),g(\lambda) $ assumed polynomials in $ \lambda $.
In \cite{AFNV-99}, the defining property of a classical ensemble in the random
matrix context was identified as the operator 
\begin{equation}
  {\bf n} = f{d \over dx} + \left({f'-g \over 2}\right) \ ,
\label{op-weight}
\end{equation}
increasing the degree of the polynomials by one. Thus for a classical ensemble
\begin{equation}
   \text{degree}\, f \leq 2, \qquad
   \text{degree}\, g \leq 1 \ .
\label{class-weight}
\end{equation}
Using this criterium one can check that the weight functions
\begin{equation}
   w_2(\lambda) = 
   \begin{cases}
      e^{-\lambda^2} \ , \qquad\text{Hermite} \\
      \lambda^a e^{-\lambda} \quad (\lambda > 0) \ , \qquad\text{Laguerre} \\
      (1\m \lambda)^{\alpha}(1\+\lambda)^{\beta}
       \quad (-1 < \lambda < 1) \ , \qquad\text{Jacobi} \\
   \end{cases}
\label{class-ops}
\end{equation}
are all classical. These weight functions all naturally occur as the eigenvalue
p.d.f. for certain ensembles of Hermitian matrices based on matrices with
independent Gaussian elements (see e.g. the Introduction section of 
\cite{FR-99}). Furthermore, for each of these cases the probability of a single
eigenvalue free region which includes an endpoint of the support of the weight
has been expressed in terms of the solution of a particular non-linear
differential equation \cite{TW-94}. The same has been done for the Hermite
and symmetric Jacobi ($ \alpha = \beta $) ensembles when the eigenvalue free 
region consists of two disjoint intervals at either end of the support of
the spectrum \cite{WFC-00}. The Cauchy weight function (\ref{CyUE-weight})
is significant for the feature of satisfying the criterium (\ref{class-weight})
and thus being a fourth classical weight function. The orthogonal polynomial
system defined by the Cauchy weight has been investigated in \cite{R-1929} and
\cite{A-1989}, where the polynomials have been found to be given by certain
Jacobi polynomials of pure imaginary argument. Consequently we find some 
similarities with the Jacobi ensemble, but also sufficient differences to
warrant separate attention.
With $E_{2}(0;I;w_2(\lambda);N)$ denoting the probability that there are no
eigenvalues in the interval $I$ of an ensemble with eigenvalue
p.d.f.~(\ref{CyUE-pdf}), the probabilities to be calculated are
\begin{align}
  & E_{2}(0;(s,\infty); (1\+\lambda^2)^{-N-a};N),
  \label{CyUE-gap:a} \\
\intertext{and}
  & E_{2}(0;(-\infty,-s) \cup (s,\infty); (1\+\lambda^2)^{-N-a};N) \ .
  \label{CyUE-gap:b}
\end{align}
The quantity (\ref{CyUE-gap:a}) gives the probability that there are no 
eigenvalues in the CyUE greater than $s$, and the quantity in 
(\ref{CyUE-gap:b}) gives the probability that there are no eigenvalues in 
the CyUE with modulus greater than $s$. In terms of the circular Jacobi
ensemble the probabilities corresponding to (\ref{CyUE-gap:a}) and 
(\ref{CyUE-gap:b}) are
\begin{align}
  & E_{2}(0;(0,x); |1\m z|^{2a};N),
  \label{cJUE-gap:a} \\
\intertext{and}
  & E_{2}(0;(-x,x); |1\m z|^{2a};N) \ .
  \label{cJUE-gap:b}
\end{align}

The details of the Cauchy ensemble are defined in Section 2, including
the orthonormal polynomials and associated coefficients. In Section 3 the 
formalism of Tracy and Widom \cite{TW-94} giving coupled differential 
equations for the gap probability and some auxiliary quantities is revised. 
In Section 4 the coupled equations for the single interval are reduced to an
ordinary differential equation specifying the probability (\ref{CyUE-gap:a}), 
and this equation is shown to be related to a \mbox{Painlev\'e-VI} transcendent.
This problem and the associated ODE are mapped back to the circular Jacobi
ensemble and the thermodynamic limit taken of (\ref{cJUE-gap:a}), and the 
resulting ODE and its transformation to a \mbox{Painlev\'e-V} type studied.
A similar study is undertaken in Section 5 for the probabilities of exclusion
from the double interval, namely (\ref{CyUE-gap:b}) and the scaling limit of
(\ref{cJUE-gap:b}).

\vfill\eject

\setcounter{equation}{0}
\section{The Cauchy Ensemble and Orthogonal Polynomials}

Consider the problems of computing the probability that there are no
eigenvalues in a region $ I $ of the spectrum for an ensemble specified by
(\ref{CyUE-weight}). Suppose $ I $ consists of $ M $ disjoint intervals, so
that with the endpoints of these intervals denoted by $ \{a_{j}\}^{2M}_{j=1} $,
\begin{equation}
  I = \bigcup^{M}_{m \geq 1} (a_{2m-1},a_{2m}) \ .
\label{interval}
\end{equation}
The probability of no eigenvalues being found in this interval is given by the
general expression (see e.g. \cite{F-00}) 
\begin{equation}
   E(0;I) = 1 + \sum^{\infty}_{n=1} {(-1)^n \over n!}
    \int_{I} dx_{1} \ldots \int_{I} dx_{n}
    \rho_{n}(x_{1},\ldots ,x_{n}) \ ,
\label{prob-0}
\end{equation}
where $ \rho_{n} $ is the $ n $-point distribution function of the eigenvalue
p.d.f. For Hermitian matrix ensembles with unitary symmetry the eigenvalue 
p.d.f. is proportional to (\ref{CyUE-pdf}), and the corresponding 
$ n $-point distribution function is given in terms of the orthonormal 
polynomials $ \{p_{j}(x)\}_{j = 0,1,2,\ldots} $ associated with the weight 
function $ w_2(x) $ according to the formula
\begin{equation}
  \rho_{n}(x_{1},\ldots ,x_{n}) = 
    \det \left[ K_N(x_{i},x_{j}) \right]_{1\leq i,j \leq n} \ ,
\label{n-particle}
\end{equation}
where 
\begin{equation}
  K_{N}(x,y) = \left[ w_2(x)w_2(y) \right]^{1/2}
  \sum^{N-1}_{l=0} p_{l}(x)p_{l}(y) \ .
\label{kernel-sum}
\end{equation}
Substituting into (\ref{prob-0}), the Fredholm theory of integral operators
then gives
\begin{equation}
  E(0;I) = \det( \Bbb{I}-\Bbb{K}_{N} ) \ ,
\label{prob-fd}
\end{equation}
where $ \Bbb{K}_{N} $ is the integral operator with kernel $ K_{N}(x,y) $ 
defined on the interval $ I $. A crucial point is that (\ref{kernel-sum}) can
be summed according to the Christoffel-Darboux formula and so written in the
special form \cite{IIKS-90}
\begin{equation}
   K_{N}(x,y) = { \phi(x)\psi(y) - \phi(y)\psi(x) \over x-y } \ ,
\label{kernel}
\end{equation}
where with $ a_{N} $ denoting the coefficient of $ x^N $ in $ p_{N}(x) $
\begin{equation}
 \begin{split}
    \phi(x) & = 
    \left( {a_{N-1} \over a_{N}}w_2(x) \right)^{1/2} p_{N}(x) \ ,
    \\
    \psi(x) & = 
    \left( {a_{N-1} \over a_{N}}w_2(x) \right)^{1/2} p_{N-1}(x) \ .
 \end{split}
\label{phi-psi}
\end{equation}

Now, the Cauchy weight is defined by (\ref{CyUE-weight}) where 
$ N \in \Bbb{Z}^+ $ and $ a \in \Bbb{R}^+ $. This weight defines an 
orthogonal polynomial system only up to the degree $ n = N $ and not for 
higher degrees, however this is of no consequence as we only require those 
below $ N $. The Cauchy weight is closely related to the symmetric Jacobi
weight $ (1\m \lambda^2)^{\alpha} $, as can be seen from the following
integration identity \cite{F-00}
\begin{lemma}
For analytic functions $ f(x) $, such that the following integrals exist
and $ \Re(\alpha) > -1 $,
\begin{equation}
   \int^{+\infty}_{-\infty} dx f(ix)(1\+ x^2)^{\alpha} =
   \tan(\pi \alpha)\int^{1}_{-1} dx f(x) (1\m x^2)^{\alpha} \ .
\label{int-identity}
\end{equation}
\end{lemma}
Using (\ref{int-identity}), the fact that
\begin{equation}
   \int^{1}_{-1} dx (1\m x^2)^{\alpha}
    P^{(\alpha,\alpha)}_m(x) P^{(\alpha,\alpha)}_n(x) = 
    {2^{2\alpha +1}\Gamma^2(\alpha\+ n\+ 1) \over 
     (2\alpha\+ 2n\+ 1)\Gamma(n\+ 1)\Gamma(2\alpha\+ n\+ 1)} \delta_{m,n} \ ,
\end{equation}
with $ P^{(\alpha,\beta)}_n(x) $ denoting the usual Jacobi polynomial,
allows us to conclude that
\begin{equation}
 p^{{\rm Cy}}_n(x) = i^n 2^{N+a}
          \left[ {n!(N\+ a\m n\m \half)\Gamma^2(N\+ a\m n)
                 \over 2\pi \Gamma(2N\+ 2a\m n)}
          \right]^{1/2} P^{(-N-a,-N-a)}_{n}(ix) \ ,
\label{cauchy-poly}
\end{equation}
are the orthonormal polynomials for the Cauchy weight. From the definition of
the $ P^{(\alpha,\beta)}_n(x) $ it follows from (\ref{cauchy-poly}) that the
coefficient of $ x^n $ required in (\ref{phi-psi}) is such that
\begin{equation}
  {a^{{\rm Cy}}_{n-1} \over a^{{\rm Cy}}_{n}} = \half
   \left[ { n(2N\+ 2a\m n) \over 
            (N\+ a\m n\+ \half)(N\+ a\m n\m \half) } \right]^{1/2} \ .
\label{cauchy-coeff}
\end{equation}
A common feature of the classical orthogonal polynomials is that $ \phi(x) $ 
and $ \psi(x) $ satisfy the recurrence-differential relations
\begin{equation}
 \begin{split}
  m(x)\phi'(x)
  & = A(x)\phi(x)+B(x)\psi(x) \ , \\
  m(x)\psi'(x)
  & = -C(x)\phi(x)-A(x)\psi(x) \ ,
\end{split}
\label{rde}
\end{equation}
where the coefficient functions $ m(x), A(x), B(x), C(x) $ are polynomials in 
$ x $.  Making use of the differentiation formula
\begin{multline}
  (2n\+ \alpha\+ \beta)(1\m x^2){d \over dx}P^{(\alpha,\beta)}_{n}(x)
  = n[\alpha\m \beta\m (2n\+ \alpha\+ \beta)x]P^{(\alpha,\beta)}_{n}(x)
  \\
  + 2(n\+ \alpha)(n\+ \beta)P^{(\alpha,\beta)}_{n-1}(x) \ ,
\label{jacobi-diff}
\end{multline}
and the three term recurrence for the Jacobi polynomials gives that (\ref{rde})
holds for the Cauchy ensemble with \cite{F-00}
\begin{equation}
 \begin{split}
  m(x) & = 1\+ x^2 \ , \\
  A(x) & = -ax \ , \\
  B(x) & = \left[ N(N\+ 2a){2a\m 1 \over 2a\+ 1} \right]^{1/2} = \beta_0 \ , \\
  C(x) & = \left[ N(N\+ 2a){2a\+ 1 \over 2a\m 1} \right]^{1/2} = \gamma_0 \ ,
\end{split}
\label{cauchy-param}
\end{equation}
with $ n = N $.

Tracy and Widom \cite{TW-94} have developed a formalism based on the integral
operator determinant formula (\ref{prob-fd}), and which makes essential use 
of the special structures (\ref{kernel}) and (\ref{rde}), to derive a set of 
coupled differential equations for $ E_2(0;I) $ and some auxiliary quantities.
The salient features of this theory will be revised in the next section.

\vfill\eject

\setcounter{equation}{0}
\section{The General Formalism}

Our goal is to characterise the probabilities (\ref{CyUE-gap:a}) and
(\ref{CyUE-gap:b}) as the solution of certain nonlinear differential 
equations. Following \cite{TW-94} this is achieved by specifying partial
differential equations for the quantities 
$ q_{j}, p_{j}, u, v, w $ and $ R(a_{j},a_{k}) $ defined henceforth.
\begin{definition}
Let $ A \doteq A(x,y) $ denote that the integral operator $ A $ has kernel
$ A(x,y) $. Then the kernels $ \rho(x,y) $ and $ R(x,y) $ are specified by
\begin{equation}
  \begin{split}
     (1-K)^{-1} 
   & \doteq \rho(x,y) \ ,
   \\
     K(1-K)^{-1} 
   & \doteq R(x,y) \ ,
  \end{split}
\label{resolvent}
\end{equation}
\end{definition}
The operator $ K(1-K)^{-1} $ is called the resolvent and $ R(x,y) $ the 
resolvent kernel.
\begin{definition}
For $ k \in \Bbb{Z}_{\geq 0} $ the functions $ Q_{k} $ and $ P_{k} $ are 
defined by
\begin{equation}
  \begin{split}
     Q_{k}(x)  
   & = \int_{I}dy\; \rho(x,y) y^{k}\phi(y) \ ,
   \\
     P_{k}(x)
   & = \int_{I}dy\; \rho(x,y) y^{k}\psi(y) \ ,
  \end{split}
\label{QP-defn}
\end{equation}
and their values at the endpoints $ a_{j} $ of $ I $ are denoted 
$ q_{kj}, p_{kj} $ so that 
\begin{equation}
  \begin{split}
     q_{kj}
   & = Q_{k}(a_{j}) \equiv \lim_{x \to a_{j}} Q_{k}(x) \ ,
   \\
     p_{kj}
   & = P_{k}(a_{j}) \equiv \lim_{x \to a_{j}} P_{k}(x) \ .
  \end{split}
\label{qp-defn}
\end{equation}
\end{definition}
Where there is no confusion, we denote $ q_{0j}, p_{0j} $ by $ q_{j}, p_{j} $.
\begin{definition}
The inner products $ u, v, w $ are defined by
\begin{equation}
  \begin{split}
     u
   & = \langle \phi | Q \rangle = \int_{I}dy\; Q_{0}(y)\phi(y) \ ,
   \\
     v
   & = \langle \psi | Q \rangle = \int_{I}dy\; Q_{0}(y)\psi(y)
     = \langle \phi | P \rangle = \int_{I}dy\; P_{0}(y)\phi(y) \ ,
   \\
     w
   & = \langle \psi | P \rangle = \int_{I}dy\; P_{0}(y)\psi(y) \ .
  \end{split}
\label{uvw-defn}
\end{equation}
\end{definition}

The coupled differential equations come in two types. There is a set of 
universal equations which are independent of the recurrence-differential 
equations (\ref{rde}), and a set of equations which depend on the details of 
(\ref{rde}). Let us first present the former.
\begin{proposition}
For general functions $ \phi(x),\psi(x) $ we have the relations
\begin{equation}
  {\partial \over \partial a_{j}} \log \det (1-K)
  = (-1)^{j-1} R(a_{j},a_{j}) \ ,
\label{diff-E}
\end{equation}
and for $  j \neq k $,
\begin{equation}
  R(a_{j},a_{k})
  = {q_{j}p_{k} - q_{k}p_{j} \over a_{j}-a_{k}} \ ,
\label{fn-R}
\end{equation}
and 
\begin{equation}
  {\partial \over \partial a_{k}} R(a_{j},a_{j})
  = (-1)^{k}R(a_{j},a_{k})R(a_{k},a_{j}) \ ,
\label{diff-R}
\end{equation}
along with
\begin{equation}
  \begin{split}
    {\partial q_{j} \over \partial a_{k}} 
   & = (-1)^{k} R(a_{j},a_{k}) q_{k}
   \ , \\
    {\partial p_{j} \over \partial a_{k}} 
   & = (-1)^{k} R(a_{j},a_{k}) p_{k}
   \ ,
  \end{split}
\label{diff-pq}
\end{equation}
for $ j \neq k $ and
\begin{equation}
  \begin{split}
    {\partial u \over \partial a_{k}}
   & = (-1)^{k} q^2_{k}
   \ , \\
    {\partial v \over \partial a_{k}}
   & = (-1)^{k} q_{k}p_{k}
   \ , \\
    {\partial w \over \partial a_{k}}
   & = (-1)^{k} p^2_{k}
   \ .
  \end{split}
\label{diff-uvw}
\end{equation}
\end{proposition}

The second set of equations, which depend on the details of (\ref{rde}),
give the $ j = k $ cases of (\ref{fn-R}) and (\ref{diff-pq}). Now for the
Cauchy weight (\ref{CyUE-weight}) we know from (\ref{cauchy-param}) the 
equations (\ref{rde}) hold for $ m(x) $ a quadratic and $ A(x), B(x), C(x) $ 
linear functions, and thus of the general form
\begin{equation}
  \begin{split}
  m(x) & = \mu_{0}+\mu_{1}x+\mu_{2}x^2 \ ,
  \\
  A(x) & = \alpha_{0}+\alpha_{1}x \ ,
  \\
  B(x) & = \beta_{0}+\beta_{1}x  \ ,
  \\
  C(x) & = \gamma_{0}+\gamma_{1}x \ .
  \end{split}
\label{poly-defn}
\end{equation}
One then has the following equations \cite{TW-94}.
\begin{proposition}
In the case that $ \phi(x), \psi(x) $ satisfy the equations (\ref{rde})
with coefficient functions (\ref{poly-defn}) we have
\begin{equation}
  \begin{split}
    m_{i}{\partial q_{i} \over \partial a_{i}}
   & = [\alpha_{0}+\alpha_{1}a_{i}+\gamma_{1}u-\beta_{1}w-\mu_{2}v] q_{i}
   \\
   & \qquad
     + [\beta_{0}+\beta_{1}a_{i}+2\alpha_{1}u+2\beta_{1}v+\mu_{2}u] p_{i}
   \\
   & \qquad\qquad
     - \sum^{2M}_{k \neq i} (-1)^k R(a_{i},a_{k}) q_{k}m_{k}
   \ , \\
    m_{i}{\partial p_{i} \over \partial a_{i}}
   & = [-\gamma_{0}-\gamma_{1}a_{i}+2\gamma_{1}v+2\alpha_{1}w-\mu_{2}w] q_{i}
   \\
   & \qquad
     + [-\alpha_{0}-\alpha_{1}a_{i}+\beta_{1}w-\gamma_{1}u+\mu_{2}v] p_{i}
   \\
   & \qquad\qquad
     - \sum^{2M}_{k \neq i} (-1)^k R(a_{i},a_{k}) p_{k}m_{k}
   \ , \\
  \end{split}
\label{diff-diag-pq}
\end{equation}
and
\begin{align}
    m_{i}R(a_{i},a_{i})
   & = [\gamma_{0}+\gamma_{1}a_{i}-2\gamma_{1}v-2\alpha_{1}w+\mu_{2}w] q^2_{i}
   \nonumber \\*
   & \qquad
     + [\beta_{0}+\beta_{1}a_{i}+2\alpha_{1}u+2\beta_{1}v+\mu_{2}u] p^2_{i}
   \nonumber \\*
   & \qquad\qquad
     + [\alpha_{0}+\alpha_{1}a_{i}+\gamma_{1}u-\beta_{1}w-\mu_{2}v] 2q_{i}p_{i}
   \nonumber \\*
   & \qquad\qquad\qquad
     + \sum^{2M}_{k \neq i} (-1)^k m_{k} 
       { [q_{i}p_{k}-p_{i}q_{k}]^2 \over a_{i}-a_{k} } \ ,
\label{diag-R}
\end{align}
and furthermore,
\begin{align}
    {\partial \over \partial a_{i}}\left[ m_{i}R(a_{i},a_{i}) \right]
   & = 2\alpha_{1} q_{i}p_{i} + \beta_{1} p^2_{i} + \gamma_{1} q^2_{i}
   \nonumber \\*
   & \qquad\qquad
     - \sum^{2M}_{k \neq i} (-1)^k m_{k} R^2(a_{i},a_{k}) \ ,
\label{diff-diag-R}
\end{align}
where $ m_{i} = m(a_{i}) $.
\end{proposition}

\vfill\eject

\setcounter{equation}{0}
\section{The Single Interval}

We first consider the probability for the interval $ (s,\infty) $ to be free
of eigenvalues so that $ a_1 = s $ and $ a_2 = \infty $. 
We shall adopt the conventions 
$ q_1,p_1 = q,p $ and $ R = R(s,s) $, noting that $ q_2,p_2 = 0 $.
\begin{proposition}
The coupled differential equations for the finite $ N $ CyUE on the interval 
$ (s,\infty) $ for general $ \alpha, \beta $ are
\begin{align}
  [\ln E_{2}]'
  & = R \ ,
  \label{Scauchy-sde:a}\\
  u'
  & = -q^2 \ ,
  \label{Scauchy-sde:b}\\
  v'
  & = -qp \ ,
  \label{Scauchy-sde:c}\\
  w'
  & = -p^2 \ ,
  \label{Scauchy-sde:d}\\
  (1\+ s^2)q'
  & = - [a s+v] q
      + [\beta_0-u(2a\m 1)] p \ ,
  \label{Scauchy-sde:e}\\
  (1\+ s^2)p'
  & = - [\gamma_0+w(2a\+ 1)] q
      + [a s+v] p \ ,
  \label{Scauchy-sde:f}\\
  (1\+ s^2)R
  & = [\gamma_0+w(2a\+ 1)] q^2
     + [\beta_0-u(2a\m 1)] p^2
     - [a s+v] 2qp
  \label{Scauchy-sde:g}\\
  \left[(1\+ s^2)R\right]'
  & =  -2a qp \ ,
  \label{Scauchy-sde:h}
\end{align}
\end{proposition}
Proof (sketch) - The first equation follows from (\ref{diff-E}) and 
(\ref{prob-fd}). The next three (\ref{Scauchy-sde:b})-(\ref{Scauchy-sde:d})
follow from (\ref{diff-uvw}) and the pair (\ref{Scauchy-sde:e}),
(\ref{Scauchy-sde:f}) follow from (\ref{diff-diag-pq}). Equation 
(\ref{Scauchy-sde:g}) follows from (\ref{diag-R}), while (\ref{Scauchy-sde:h})
follows from (\ref{diff-diag-R}). 
$ \square $

The boundary conditions satisfied by $ R(s,s) $ as $ s \to \infty $, deduced
from the fact that in this limit $ R(s,s) \sim K_N(s,s) $, can be expressed as
\begin{align}
  R(s,s) \sim \,
  &  (-1)^{N-1}{ 2^{2(N+a)} \over 4\pi } 
     { N!\Gamma^2(a) \over \Gamma(N\+ 2a)}(1\+ s^2)^{-N-a-1}
    \nonumber \\*
  & \qquad \times
     \left\{- N(N\+ 2a)\left[ P^{(-N-a,-N-a)}_{N}(is) \right]^2 
     \right.
    \nonumber \\*
  & \qquad\phantom{\times\Biggl\}}
     - 2ia^2s P^{(-N-a,-N-a)}_{N}(is)P^{(-N-a,-N-a)}_{N-1}(is)
    \nonumber \\*
  & \qquad\phantom{\times\Biggl\}}
     \left.
     + a^2\left[ P^{(-N-a,-N-a)}_{N-1}(is) \right]^2 \right\} \ ,
\label{Scauchy-bc1}
\end{align}
or considering only the first two leading order terms
\begin{align}
  (1\+ s^2) R(s,s) \sim
  &  { 2^{2a} \over \pi } 
     { \Gamma(1\+ N\+ 2a)\Gamma^2(a\+ 1) \over \Gamma(N)\Gamma^2(2a\+ 2)}
     s^{-2a} \nonumber \\*
  & \qquad \times
     \left[ 2a\+ 1 
            - s^{-2}{a \over 2a\+ 3}(2N^2\+ 4aN\+ 4a^2\+ 4a\+ 1) \right] \ .
\label{Scauchy-bc2}
\end{align}

We now indicate how to reduce such a system to a single second order 
differential equation for $ R = R(s) = R(s,s) $.
\begin{proposition}
The coupled set of ODEs given in Proposition 3 reduce to the second order 
ODE for $ \sigma(s) = (1\+ s^2)R(s) $,
\begin{align}
 & (1\+ s^2)^2(\sigma'')^2 + 4(1\+ s^2)(\sigma')^3
       - 8s\sigma(\sigma')^2 + 4\sigma^2(\sigma'-a^2)
 \nonumber \\*
 & \qquad
   + 8a^2s\sigma\sigma'
   + 4\left[ N(N\+ 2a) - a^2s^2 \right](\sigma')^2 = 0 \ .
\label{Scauchy-ode2}
\end{align}
\end{proposition}
Proof - We note from (\ref{Scauchy-sde:c}) and (\ref{Scauchy-sde:h}) the 
integral
\begin{equation}
  (1\+ s^2)R = 2a v \ ,
\label{Scauchy-int1}
\end{equation}
where the constant of integration has been set to zero because
$ v, (1\+ s^2)R \to 0 $ as $ s \to \infty $.
Now we show how a second integral of the motion can be found. Utilising this
first integral and (\ref{Scauchy-sde:g}) we have
\begin{equation}
   [\gamma_0+w(2a\+ 1)] q^2
 + [\beta_0-u(2a\m 1)] p^2
 - [a s+v] 2qp - 2av = 0 \ ,
\label{Scauchy-aux1}
\end{equation}
while adding $ p $ times (\ref{Scauchy-sde:e}) and $ q $ times 
(\ref{Scauchy-sde:f}) yields another equation quadratic in the $ p $ and $ q $
\begin{equation}
  (1\+ s^2)(pq)'
 + [\gamma_0+w(2a\+ 1)] q^2
 - [\beta_0-u(2a\m 1)] p^2 = 0 \ .
\label{Scauchy-aux2}
\end{equation}
Now the idea is to combine these two equations in such a way so that use of the
relations (\ref{Scauchy-sde:b}-\ref{Scauchy-sde:d}) to express products of
$ q, p $ in terms of derivatives of $ u, v, w $ leads to an exact derivative.
Such a combination is 
(\ref{Scauchy-aux1}) minus $ 2a $ times (\ref{Scauchy-aux2}), which when
equations (\ref{Scauchy-sde:b}-\ref{Scauchy-sde:d}) are employed yields
\begin{align}
 & 2a(1\+ s^2)v'' + 2asv' + 2vv' - 2av
 \nonumber \\*
 & + (2a\m 1)u'[\gamma_0+w(2a\+ 1)] - (2a\+ 1)w'[\beta_0-u(2a\m 1)] = 0 \ .
\label{Scauchy-comb1}
\end{align}
This can be rewritten as 
\begin{align}
 & 2a(1\+ s^2)v'' + 4asv'
 \nonumber \\*
 & + 2vv' - 2av - 2asv'
 \nonumber \\*
 & - [\beta_0-u(2a\m 1)]'[\gamma_0+w(2a\+ 1)]
   - [\gamma_0+w(2a\+ 1)]'[\beta_0-u(2a\m 1)] = 0 \ ,
\label{Scauchy-comb2}
\end{align}
which is a perfect derivative. Given that the boundary conditions as
$ s \to \infty  $ imply $ u, v, w \to 0 $, the integration constant can be
determined, leading to the second integral
\begin{equation}
  [\beta_0-u(2a\m 1)][\gamma_0+w(2a\+ 1)] = 
  \beta_0\gamma_0 + (1\+ s^2)\sigma' - s\sigma + {1\over 4a^2}\sigma^2 \ .
\label{Scauchy-int2}
\end{equation}
We seek now to utilise all these equations to arrive at a second order ODE
for $ \sigma(s) $, and to do so we note the pair of equations
\begin{equation}
  - {1\over 2a}(1\+ s^2)\sigma'' =
    [\beta_0-u(2a\m 1)]p^2 - [\gamma_0+w(2a\+ 1)]q^2 \ ,
\label{Scauchy-elim1}
\end{equation}
which is a consequence of (\ref{Scauchy-aux2}) and (\ref{Scauchy-int1}),
and
\begin{equation}
  \sigma - {1\over 2a^2}(\sigma + 2a^2s)\sigma' = 
  [\beta_0-u(2a\m 1)]p^2 + [\gamma_0+w(2a\+ 1)]q^2 \ .
\label{Scauchy-elim2}
\end{equation}
We seek to eliminate the two terms appearing on the right-hand sides of these
last equations by forming the difference of their squares and using their
cross product from the second integral (\ref{Scauchy-int2}). The resulting
equation is the ODE stated in the proposition.
$ \square $

\subsection{Reduction to \mbox{Painlev\'e} Type}
It is immediately clear that we can make an identification of our ODE with
the master \mbox{Painlev\'e} equation for second order second degree ordinary
differential equations with \mbox{Painlev\'e} integrability of Cosgrove and 
Scoufis, \cite{CS-93} equation (5.1), where in terms of this equation we have
$ g(s) = 1\+ s^2, \; h(s) = N(N\+ 2a)-a^2s^2, \; f(s) = 0 $, 
and the coefficients 
$ c_1 = 0, c_2 = 1, c_3 = 0, c_4 = 1, c_5 = -a^2, 
  c_6 = 0, c_7 = N(N\+ 2a), c_8 = 0, c_9 = 0, c_{10} = 0 $. 
\begin{proposition}
The function $ \sigma(s) = (1\+ s^2)R(s) $ for the Cauchy Unitary Ensemble
on the interval $ (s,\infty) $ is related to a solution of the
\mbox{Painlev\'e} VI transcendent $ \omega(t) $ with the parameters
\begin{equation} 
  \alpha = \half           \ ,\qquad
  \beta  = -\half(N\+ a)^2  \ ,\qquad
  \gamma =  \half(N\+ a)^2 \ ,\qquad
  \delta = \half - 2a^2 \ ,
\label{Scauchy-Pparam}
\end{equation} 
by the gauge 
\begin{equation}
\begin{split}
       t   & = \half(1\m is) \ ,\\
  \eta(t)  & = {1\over 2i}(\sigma(s)-a^2s) \ ,
\end{split}
\label{Scauchy-Pgauge}
\end{equation}
and B\"acklund transformations
\begin{equation}
\begin{split}
  \eta  & =
  {t^2(t\m 1)^2 \over 4\om(\om\m 1)(\om\m t)}
  \left\{ \dot{\om} - {\om(\om\m 1) \over t(t\m 1)} \right\}^2
  - \tfrac{1}{4}(N\+ a)^2{\om\m t \over \om(\om\m 1)}
  + \half a^2\left( 1 - 2t{\om\m 1 \over \om\m t}\right) \ ,
  \\
  \dot{\eta}  & =
  - {t(t\m 1) \over 4\om(\om\m 1)}
  \left\{ \dot{\om} - {\om(\om\m 1) \over t(t\m 1)} \right\}^2
  + \tfrac{1}{4}(N\+ a)^2{(\om\m t)^2 \over t(t\m 1)\om(\om\m 1)} \ .
\end{split}                        
\label{Scauchy-Pbaecklund}
\end{equation}
\end{proposition}
All that is required to make an explicit relation to one of the 
\mbox{Painlev\'e} transcendents is to find the right gauge transformation 
to put (\ref{Scauchy-ode2}) into the canonical form of (SD-Ia) in \cite{CS-93}.
Given that our function $ g(s) $ has zeros at 
$ s_1 = \infty, s_2 = i, s_3 = -i $ this determines $ a_1, a_2, a_3, a_4 $
in the general gauge transformation
\begin{equation}
  t = {a_1s + a_2 \over a_3s + a_4} \qquad 
  \eta = {a_5\sigma + a_6s + a_7 \over a_3s + a_4} \ ,
\end{equation}
leaving $ a_5, a_6, a_7 $. The coefficient $ a_5 $ is found from matching
the cubic terms in the ODE (all three cubic terms must be matched with the
single coefficient), whilst $ a_6, a_7 $ are determined by the
$ \eta^2, \eta\dot{\eta} $ terms. 
This determines the gauge transformation as given in (\ref{Scauchy-Pgauge}).
The free parameters in the canonical form are then found to be
\begin{equation}
\begin{split}
   A_1 & = 3a^2+N(N\+ 2a) \ ,\\
   A_2 & =  0              \ ,\\
   A_3 & =  3a^4+2a^2N(N\+ 2a) \ ,\\
   A_4 & = a^6+a^4N(N\+ 2a) \ .
\end{split}
\label{Scauchy-Aparam}
\end{equation}
From these four coefficients one solution for the \mbox{Painlev\'e-VI}
transcendent coefficients is given in (\ref{Scauchy-Pparam}) and the 
corresponding B\"acklund transformation by (\ref{Scauchy-Pbaecklund}).
Other solution sets are possible, which are manifestations of Schlesinger 
transformations, and all are tabulated below in Table 1.
$ \square $

\renewcommand{\arraystretch}{1.5}
\begin{table}

\begin{tabular}{|c|c|c|c|}
\hline
 $ \alpha = $ & $ \beta = $ & $ \gamma = $ & $ \delta = $ \\
\hline
  $ \half $ 
& $ -\half(N\+ a)^2 $ 
& $ \half(N\+ a)^2 $ 
& $ \half - 2a^2 $ \\
\hline

  $ \half(1\pm 2a)^2 $ 
& $ -\half(N\+ a)^2 $ 
& $ \half(N\+ a)^2 $ 
& $ \half $ \\
\hline

  $ \half[1\pm (N\+ 2a)]^2 $ 
& $ -\half a^2 $ 
& $ \half a^2 $ 
& $ \half(1-N^2) $ \\
\hline

  $ \half[1\pm N]^2 $ 
& $ -\half a^2 $ 
& $ \half a^2 $ 
& $ \half[1-(N\+ 2a)^2] $ \\
\hline

  $ \half[1\pm (N\+ a)]^2 $ 
& $ 0 $ 
& $ 2a^2 $ 
& $ \half[1-(N\+ a)^2] $ \\
\hline

  $ \half[1\pm (N\+ a)]^2 $ 
& $ -2a^2 $ 
& $ 0 $ 
& $ \half[1-(N\+ a)^2] $ \\
\hline

  $ \half[1\pm a]^2 $ 
& $ -\half N(N\+ 2a) $ 
& $ 2a^2+\half N(N\+ 2a) $ 
& $ \half(1- a^2) $ \\
\hline

  $ \half[1\pm a]^2 $ 
& $ 2a^2-\half N(N\+ 2a) $ 
& $ \half N(N\+ 2a) $ 
& $ \half(1- a^2) $ \\
\hline

\end{tabular}

\bigskip
\caption{
\mbox{Painl\'eve} parameters for the P-VI transcendents characterising the
gap p.d.f. for the Cauchy Hermitian random matrix ensemble on the interval 
$ (s,\infty) $.}
\label{Scauchy-table}
\bigskip

\end{table}

\subsection{Special Cases of low $ N $}
In this part we present the calculations for the first two finite-$ N $ cases,
that is $ N = 1 $ and $ N = 2 $, by direct means using the probability 
$ E_2(0;I) $.
\begin{proposition}
The probability (\ref{CyUE-gap:a}) for $ N = 1 $ and the associated quantities 
are given in terms of the Gauss hypergeometric function 
$ {}_{2}F_{1}(a,b;c;z) $ by
\begin{equation}
\begin{split}
  E_2(0;I)
  & = X \ , \\
  \sigma(s)
  & = {\Gamma(a\+ 1) \over \sqrt{\pi}\Gamma(a\+ \half)}
      {(1\+ s^2)^{-a} \over E_2(0;I)} \ , 
\end{split}
\label{Scauchy-N1}
\end{equation}
and similarly for the probability with $ N = 2 $ ,
\begin{equation}
\begin{split}
  E_2(0;I)
  & =  X^2 - {a\Gamma(a\+ 1) \over \sqrt{\pi}\Gamma(a\+ \thalf)}
             s(1\+ s^2)^{-a-1} X 
           - {\Gamma^2(a\+ 1) \over 2\pi\Gamma(a\+ \half)\Gamma(a\+ \thalf)}
             (1\+ s^2)^{-2a-1} \ , \\
  \sigma(s)
  & =  {\Gamma(a\+ 2) \over \sqrt{\pi}\Gamma(a\+ \thalf)}
       {(1\+ s^2)^{-a-1} \over E_2(0;I)}
       \left\{ [1+(2a\+ 1)s^2]X
               + {\Gamma(a\+ 1) \over \sqrt{\pi}\Gamma(a\+ \half)}
                 s(1\+ s^2)^{-a} \right\} \ ,
\end{split}
\label{Scauchy-N2}
\end{equation}
where
\begin{equation}
 \begin{split}
  X
  & = 1 - {\Gamma(a\+ 1) \over 2\sqrt{\pi}\Gamma(a\+ \thalf)}
       s^{-2a-1}\, {}_{2}F_{1}(a\+ 1,a\+ \half;a\+ \thalf;-s^{-2})  \ , \\
  & = \half + {\Gamma(a\+ 1) \over \sqrt{\pi}\Gamma(a\+ \half)}
       s\, {}_{2}F_{1}(a\+ 1,\half;\thalf;-s^{2})  \ ,
\end{split}
\label{Scauchy-X}                                                              
\end{equation}
\end{proposition}
Proof - These follow from the integral representations and transformation
formulae of the Gauss hypergeometric functions.
$ \square $

One can show that these two specific cases are solutions of the second order
differential equation (\ref{Scauchy-ode2}) and satisfy the boundary conditions 
(\ref{Scauchy-bc2}), after noting the differentiation formula for the Gauss 
hypergeometric function.

\subsection{The Thermodynamic Limit for the Circular Jacobi Ensemble}
Using the stereographic projection we map the extended real line back
onto the circle of circumference $ L $ via
\begin{equation}
  s = \cot{\pi x \over L} \quad x \in [0,L) \ .
\label{stereo}
\end{equation}
Now as 
\begin{equation}
   E_{2}(0;(s,\infty);(1\+ s^2)^{-N-a};N) = 
   \exp\left(-\int^{\infty}_{s} dt\;{\sigma(t) \over 1\+ t^2}\right) \ ,
\label{Scauchy-prob-int}
\end{equation}
this gap probability is then expressed as
\begin{equation}
   E_{2}(0;(0,x);|1\m e^{2\pi iy/L}|^{2a};N) = 
   \exp\left(-{\pi \over L}\int^{x}_{0} dy\;\sigma(\cot\pi y/L)\right) \ .
\label{Ecirc-prob-int}
\end{equation}
One is then interested in taking the scaling limit, interpreted as the 
thermodynamic limit, in which both $ N, L \to \infty $ with a fixed
density $ \rho = N/L $. Evidently the natural quantity to consider in
this limit is 
\begin{equation}
   {1 \over N}\sigma(\cot\pi x/L) \mapsto \sigma(\pi\rho x)  \ ,
\label{ETlimit-scale}
\end{equation}
and so we investigate the above limit of (\ref{Scauchy-ode2}) in terms of
this new variable.

\begin{proposition}
In the thermodynamic limit the scaled variable $ \tau_a(x) \equiv x\sigma(x) $ 
satisfies the second order second degree ordinary differential equation
\begin{equation}
  x^2(\ddot{\tau}_a)^2 - 4(x\dot{\tau}_a-\tau_a)\dot{\tau}^2_a
 - 4a^2\dot{\tau}^2_a + 4(x\dot{\tau}_a-\tau_a)^2 = 0 \ ,
\label{Escale-ode2}
\end{equation}
subject to the boundary conditions
\begin{equation}
 \tau_a(x) \underset{x \to 0}{\sim} 
  {(\half x)^{2a+1} \over \Gamma(a\+\half)\Gamma(a\+\thalf)} \ .
\label{Escale-bc}
\end{equation}
\end{proposition}
Proof - This follows directly from the definition (\ref{ETlimit-scale}) under
the map (\ref{stereo}) of (\ref{Scauchy-ode2}).
$ \square $

\begin{proposition}
The scaled variable $ \tau_a(x) $ is determined by a solution of the 
\mbox{Painlev\'e-V} transcendent $ u(x) $ with parameters
\begin{equation}
   \alpha  = \half(1\mp 2a)^2 \ ,\qquad
   \beta   = 0 \ ,\qquad
   \gamma  = 0 \ ,\qquad
   \delta  = 2 \ ,
\label{Escale-Pparam}
\end{equation}
and is related by
\begin{equation}
\begin{split}
  -\tau_a        & = 
   {1\over 4u}\left[{xu' \over u\m 1} - u\right]^2 - a^2u
   + {x^2u \over (u\m 1)^2} \ ,\\
  -\dot{\tau}_a  & = 
  -{x\over 4u(u\m 1)}\left[u' - (1\mp 2a){u(u\m 1) \over x}\right]^2
   - {xu \over u\m 1} \ .
\end{split}
\label{Escale-xfm}
\end{equation}
\end{proposition}
Proof - The equation (\ref{Escale-ode2}) falls into the class SD-I.b, equation
(5.5) of \cite{CS-93} under $ \tau_a \mapsto -\tau_a $ and so it is a matter
of finding the appropriate gauge transformation. Clearly the parameters
$ a_2 = a_3 = 0, a_1 = a_4 = a_5 = 1 $, and this leaves $ a_6, a_7 $.
Under such a mapping the canonical equation can be compared with our
(\ref{Escale-ode2}), and thus we find $ a_6 = 0, a_7 = -a^2 $ and
$ A_1 = -4, A_2 = -8a^2, A_3 = 0, A_4 = -4a^4 $. Solving for the 
\mbox{Painlev\'e-V} transcendent parameters and the transformation then leads 
to the results in the above proposition. Another valid solution has the
parameters $ \alpha = \half,\; \beta = -2a^2,\; \gamma = 0,\; \delta = 2 $.
$ \square $

In the case $ a = 0 $ the eigenvalue p.d.f. defining the cJUE is identical 
to the eigenvalue p.d.f. for the CUE of random unitary matrices. It is well
known that in the scaled thermodynamic limit the $ n-$point distribution
function then has the structure (\ref{n-particle}) with $ K_N(x,y) $
replaced by $ K(x,y) $ , which in turn has the explicit form
\begin{equation}
   K(x,y) = {\sin\pi\rho(x-y) \over \pi(x-y)} \ .
\label{sin-kernel}
\end{equation}
For this scaled ensemble it is a celebrated result of Jimbo et al 
\cite{JMMS-80} that
\begin{equation}
   E_2(0;(0,x)) = \exp\left( -\int^{\pi\rho x}_0 {\tau(y) \over y}dy \right) \ ,
\label{sin-gap}
\end{equation}
where $ \tau(x) $ satisfies the equation (\ref{Escale-ode2}) with $ a = 0 $.

We have already remarked in the Introduction that the case $ a = 1 $ of the 
cJUE is also related to the CUE, by measuring angles from a particular 
eigenvalue, taken as the origin, in the latter. Thus
\begin{equation}
   \left. E_2(0;(0,\theta);w_2(\theta)=1;N\+ 1) 
   \right|_{\text{given an eigenvalue at}\, \theta = 0}
   = E_2(0;(0,\theta);|1\m z|^2;N) \ .
\label{prob-interp}
\end{equation}
But a simple argument shows it is also true that 
\begin{equation}
   \left. {N \over 2\pi}E_2(0;(0,\theta);w_2(\theta)=1;N\+ 1) 
   \right|_{\text{given an eigenvalue at}\, \theta = 0}
   = -{d \over d\theta}E_2(0;(0,\theta);w_2(\theta)=1;N\+ 1) \ ,
\label{prob-density}
\end{equation}
where $ N/2\pi $ is the density of eigenvalues at $ \theta = 0 $, so we have 
the identity
\begin{equation}
   -{2\pi \over N}{d \over d\theta}E_2(0;(0,\theta);w_2(\theta)=1;N\+ 1)
   = E_2(0;(0,\theta);|1\m z|^2;N) \ .
\label{relation-1}
\end{equation}
Putting 
\begin{equation}
   E_2(0;(0,x);a) \equiv 
   \lim_{N \to \infty}E_2(0;(0,2\pi x/L);|1\m e^{i\theta}|^{2a};N) \ ,
\label{prob-limit}
\end{equation}
this implies
\begin{equation}
   -{1 \over \rho}{d \over dx}E_2(0;(0,x);0)
    = E_2(0;(0,x);1) 
    = \exp\left( -\int^{\pi\rho x}_0 {\tau_1(y) \over y}dy \right) \ ,
\label{Ecirc-pdf-identity}
\end{equation}
where the second equality follows from (\ref{Ecirc-prob-int}) and 
Proposition 7.

We have noted the identity (\ref{Ecirc-pdf-identity}) in a recent Letter
\cite{FW-00} presenting the exact Wigner surmise form for the spacing 
probability in the bulk of matrix ensembles with orthogonal or unitary
symmetries. This is possible (in the case of unitary symmetry) because the
spacing probability between consecutive eigenvalues, denoted $ p_2(x) $ say,
is related to $ E_2(0;(0,x);0) $ by
\begin{equation}
   p_2(x) = {d^2 \over dx^2}E_{2}(0;(0,x);0) \ ,
\label{p2-defn}
\end{equation}
where the mean spacing in normalised so that $ \rho = 1 $. Use of the formula
(\ref{Ecirc-pdf-identity}) gives \cite{FW-00}
\begin{equation}
   p_2(x) = {\tau_1(\pi x) \over x} 
            \exp\left( -\int^{\pi x}_0 {\tau_1(y) \over y}dy \right)\ ,
\label{p2-wigner}
\end{equation}
where $ \tau_1 $ satisfies the non-linear equation (\ref{Escale-ode2}) with 
$ a = 1 $ subject to the boundary condition
\begin{equation}
   \tau_1(x) \underset{x \to 0}{\sim}  {x^3 \over 3\pi} \ .
\label{tau-bc}
\end{equation}

Substituting for $ E_{2}(0;(0,x);0) $ on the left hand side of 
(\ref{Ecirc-pdf-identity}) according to (\ref{Ecirc-prob-int}) and the result
of Proposition 8 shows that (\ref{Ecirc-pdf-identity}) implies an identity
between transcendents, which in fact can be established directly from the
theory in \cite{CS-93}.
\begin{proposition}
The solutions $ \tau_0(x) $ and $ \tau_1(x) $ specified in Proposition 7
are related by
\begin{equation}
   \tau_{1}(x) = 1 + \tau_{0}(x) - x{\tau_{0}'(x) \over \tau_{0}(x)} \ .
\label{Ecirc-tau-identity}
\end{equation}
\end{proposition}
Proof - In Proposition 8 for the general B\"acklund transformations, we have 
in the case $ a = 0 $
\begin{align}
  -\tau_{0}        & =
   {1\over 4u_{0}}\left[{xu'_{0} \over u_{0}\m 1} - u_{0}\right]^2
   + {x^2u_{0} \over (u_{0}\m 1)^2}
   \ ,\nonumber\\
  -\tau'_{0}  & =
   - {x\over 4u_{0}(u_{0}\m 1)}\left[u'_{0}
   - {u_{0}(u_{0}\m 1) \over x}\right]^2
   - {xu_{0} \over u_{0}\m 1} \ .
\label{xfm-pair-0}
\end{align}
so that if we form the quantity on the right-hand side of 
(\ref{Ecirc-tau-identity}) then we find
\begin{equation}
  1 + \tau_{0}(x) - x{\tau'_{0}(x) \over \tau_{0}(x)} =
  - {1\over 4u_{0}}\left[{xu'_{0} \over u_{0}\m 1} - u_{0}\right]^2 
  + u_{0} - {x^2u_{0} \over (u_{0}\m 1)^2} \ .
\end{equation}
But now we note that the solution of the \mbox{Painlev\'e} transcendent for 
$ a = 0 $, $ u_{0} $, corresponds to $ \alpha = 1/2 $, but is also identical 
to the one for $ a = 1 $, $ u_{1} $, because $ \alpha $ is the same (taking 
the negative sign in (\ref{Escale-Pparam})) and thus one can make the 
identification
\begin{equation}
  - \tau_{1}(x) =
    {1\over 4u_{0}}\left[{xu'_{0} \over u_{0}\m 1} - u_{0}\right]^2
  - u_{0} + {x^2u_{0} \over (u_{0}\m 1)^2} \ ,
\end{equation}
so that (\ref{Ecirc-tau-identity}) is established.
$ \square $

To understand that (\ref{p2-wigner}) is the simplest possible way of 
representing the spacing probability one only needs to observe that the 
identification of the solutions of two distinct second order second degree 
ODEs with one transcendent can only arise for $ a = 0, 1 $ in 
$ \alpha = \half(1-2a)^2 $.

\vfill\eject

\setcounter{equation}{0}
\section{The Double Interval}

We will now consider our second case, which is the probability of eigenvalues
being excluded from the interval $ (-\infty,-s) \cup (s,\infty) $,
in a parallel manner to that of the previous case. 
We make the new conventions $ a_1 = -\infty, a_2 = -s, a_3 = s, a_4 = \infty $,
and $ q_3, p_3 = q, p $, with the symmetries 
$ q_2 = (-1)^{N}q, p_2 = (-1)^{N-1}p $. 
We note the parity relation implies $ v = 0 $. 
\begin{proposition}
The coupled differential equations for the finite $ N $ CyUE on the interval 
$ (-\infty,-s) \cup (s,\infty) $ are
\begin{align}
  [\ln E_{2}]'
  & = 2R \ ,
  \label{Dcauchy-sde:a}\\
  R(-s,s)
  & = (-1)^{N-1}{ qp \over s } \equiv (-1)^{N-1}R_{0}(s) \ ,
  \label{Dcauchy-sde:b}\\
  u'
  & = -2q^2 \ ,
  \label{Dcauchy-sde:c}\\
  w'
  & = -2p^2 \ ,
  \label{Dcauchy-sde:d}\\
  (1\+ s^2)q'
  & = - a sq
      + [\beta_0-u(2a\m 1)] p
      + {2(1\+ s^2) \over s}q^2p \ ,
  \label{Dcauchy-sde:e}\\
  (1\+ s^2)p'
  & = + a sp
      - [\gamma_0+w(2a\+ 1)] q
      - {2(1\+ s^2) \over s}qp^2 \ ,
  \label{Dcauchy-sde:f}\\
  (1\+ s^2)R
  & = [\gamma_0+w(2a\+ 1)] q^2
     + [\beta_0-u(2a\m 1)] p^2
     - 2a sqp
     + 2s(1\+ s^2)R^2_{0}  \ ,
  \label{Dcauchy-sde:g}\\
  \left[(1\+ s^2)R\right]'
  & = - 2a qp - {2(1\+ s^2) \over s^2}q^2p^2 \ .
  \label{Dcauchy-sde:h}
\end{align}
\end{proposition}
Proof - These follow in an entirely parallel manner as for the single interval
case.
$ \square $

Again such a system can be reduced to single second order differential 
equations for $ R(s) $ or $ R_{0}(s) $.
\begin{proposition}
The coupled set of ODEs given in Proposition 10 are equivalent to the following
second order ODE for $ \sigma(s) = (1\+ s^2)R(s) $,
\begin{align}
 & \left\{ (1\m s^2)F-2as
        + {s(1\+ s^2) \over F}\left[
          a^2s - 2s\sigma' - (1\+ s^2)\sigma'' \right]
   \right\}^2
 \nonumber \\*
 & \qquad
   - \biggl\{ 2(1\+ s^2)\sigma + 2as^2(F-as) - s(F-as)^2 \biggr\}^2
 \nonumber \\*
 & \qquad
   = -4s^2(F-as)^2 \left\{ N(N\+ 2a) -2s\sigma -2as(F-as) \right\} \ ,
\label{Dcauchy-ode2a}
\end{align}
where $ F \equiv \sqrt{a^2 s^2 - 2(1\+ s^2)\sigma'} $, with the boundary
conditions
\begin{equation}
   \sigma(s) \underset{s \to \infty}{\sim}
   {1 \over 2^{2a+1}\Gamma(a\+\half)\Gamma(a\+\thalf)}
   {\Gamma(N\+ 2a\+ 1) \over \Gamma(N)} s^{-2a}  \ .
\label{Dcauchy-bca}
\end{equation}
Alternatively the coupled ODEs can be reduced to the following second order
ODE for $ R_{0}(s) $,
\begin{align}
 & \left\{ s(1\+ s^2)^2R''_{0} + 2(1\+ s^2)(1\+ 2s^2)R'_{0}
           + T\left[ 2N(N\+ 2a) + 1\+ s^2 \right] - 6aT^2 - 4T^3
   \right\}^2
 \nonumber \\*
 & \qquad
   = \left[ as - 2(1\m s^2)R_{0} \right]^2
     \left\{ (1\+ s^2)^2 (T')^2 - 4T^2 \left[ T(T\+ 2a) - N(N\+ 2a) \right]
     \right\} \ .
\label{Dcauchy-ode2b}
\end{align}
where $ T(s) \equiv 2sR_{0} = 2qp $, subject to the boundary conditions
\begin{equation}
   R_{0}(s) \underset{s \to \infty}{\sim}
   {1 \over 2^{2a+1}\Gamma(a\+\half)\Gamma(a\+\thalf)}
   {\Gamma(N\+ 2a\+ 1) \over \Gamma(N)} s^{-2a-2} \ .
\label{Dcauchy-bcb}
\end{equation}
\end{proposition}
Proof - In order to find our second order ODE we require an integral of the
motion (not two as for the single interval case because $ v = 0 $), and
the construction of this parallels the earlier case. Firstly we note 
\begin{equation}
  (1\+ s^2)(qp)' + [\gamma_0+w(2a\+ 1)] q^2 - [\beta_0-u(2a\m 1)] p^2 = 0 \ ,
\label{Dcauchy-aux1}
\end{equation}
using (\ref{Dcauchy-sde:e},\ref{Dcauchy-sde:f}) and
\begin{equation}
  \sigma - [\gamma_0+w(2a\+ 1)] q^2 - [\beta_0-u(2a\m 1)] p^2
     + 2a sqp - 2s^{-1}(1\+ s^2)q^2p^2 = 0 \ ,
\label{Dcauchy-aux2}
\end{equation}
which is (\ref{Dcauchy-sde:g}) rewritten. Now we employ a certain combination
of these last two equations, namely (\ref{Dcauchy-aux2}) + $ 2a $ times
(\ref{Dcauchy-aux1}), which upon the use of 
(\ref{Dcauchy-sde:c},\ref{Dcauchy-sde:d}) becomes
\begin{align}
 & \half \left\{[\gamma_0+w(2a\+ 1)][\beta_0-u(2a\m 1)] \right\}'
 \nonumber \\
 & \quad
   + 2a(1\+ s^2)(qp)' + 4asqp
 \nonumber \\*
 & \qquad\qquad
   + \sigma - 2asqp - 2s^{-1}(1\+ s^2)q^2p^2 = 0 \ .
\label{Dcauchy-elim}
\end{align}
However the last two terms of this equation are just $ s\sigma' $ and thus
the whole left-hand side is a perfect derivative. Utilising the boundary
conditions as $ s \to \infty $ we arrive at the integral of motion,
\begin{equation}
   [\gamma_0+w(2a\+ 1)][\beta_0-u(2a\m 1)] =
   \beta_0\gamma_0 - 2s\sigma -4a(1\+ s^2)qp \ .
\label{Dcauchy-int}
\end{equation}
As our starting point in the reduction to an ODE in terms of $ \sigma $
we consider the two equations, (\ref{Dcauchy-aux1},\ref{Dcauchy-aux2}),
rewritten in the following way
\begin{align}
   (1\+ s^2)(qp)' 
  & =  [\beta_0-u(2a\m 1)] p^2 - [\gamma_0+w(2a\+ 1)] q^2 \ ,
  \nonumber\\
  \sigma + 2a sqp - 2s^{-1}(1\+ s^2)q^2p^2
  & = [\beta_0-u(2a\m 1)] p^2 + [\gamma_0+w(2a\+ 1)] q^2 \ .
\label{Dcauchy-start}
\end{align}
Now we view (\ref{Dcauchy-sde:h}) as a quadratic equation for $ qp $, with
\begin{equation}
   qp = {s \over 2(1\+ s^2)}[F-as] \ ,
\end{equation} 
and use this relation and the definition of $ F $ to express the left-hand
sides of (\ref{Dcauchy-start}) solely in terms of $ F $ and $ \sigma $.
Finally the difference of the squares of each are taken, and the integral
of the motion (\ref{Dcauchy-int}) is used to eliminate the cross term.
One is then left with the first of the second order ODEs (\ref{Dcauchy-ode2a}).

For the second ODE we seek to first eliminate $ [\gamma_0+w(2a\+ 1)] q^2 $ and
$ [\beta_0-u(2a\m 1)] p^2 $ by employing (\ref{Dcauchy-sde:g}) (the sum of
these two terms), (\ref{Dcauchy-aux1}) (the difference of the two) and 
(\ref{Dcauchy-int}) (the cross product of the two). In this way we find the
following quadratic equation for  $ \sigma $
\begin{equation}
   4\biggl\{ \sigma + T[as - (1\m s^2)R_{0}] \biggr\}^2 
   + 4T^2\biggl\{ T(T\+ 2a) - N(N\+ 2a) \biggr\} - (1\+ s^2)^2{T'}^2 = 0 \ ,
\label{Dcauchy-quad}
\end{equation}
and can then use this equation in conjunction with (\ref{Dcauchy-sde:h})
to eliminate $ \sigma $ entirely. The final result is the second ODE
(\ref{Dcauchy-ode2b}).
$ \square $

Remark - Knowledge of the above integral (\ref{Dcauchy-int}) allows us to
express the auxiliary quantities $ q, p, u, w $ in terms of the primary 
ones $ \sigma, R_0 $, and this will facilitate further understanding of the 
relationship of these to the transcendents and their scaling behaviour. 
\begin{proposition}
The auxiliary quantities $ q, p $ are determined by quadrature from the two
primary quantities $ \sigma, R_0 $ by the differential equations
\begin{equation}
\begin{split}
  (1\+ s^2) {q' \over q}
  & = {\sigma \over 2sR_0} + (1\+ s^2)R_0
  \\
  & \qquad
    - \left\{ \left[ {\sigma \over 2sR_0}+as-(1\+s^2)R_0 \right]^2
            +2s\sigma+4as(1\+ s^2)R_0-\beta_0\gamma_0 \right\}^{1/2} \ ,
  \\
  (1\+ s^2) {p' \over p}
  & = -{\sigma \over 2sR_0} - (1\+ s^2)R_0
  \\
  & \qquad
    - \left\{ \left[ {\sigma \over 2sR_0}+as-(1\+s^2)R_0 \right]^2
            +2s\sigma+4as(1\+ s^2)R_0-\beta_0\gamma_0 \right\}^{1/2} \ ,
\end{split}
\label{Dcauchy-pq}
\end{equation}
subject to the boundary conditions
\begin{equation}
\begin{split}
   q(s)
  & \underset{s \to \infty}{\sim}
   {1\over \Gamma(a\+\half)} \left({2a\m 1 \over 2a\+ 1}\right)^{1/4}
   \left\{ \sqrt{N(N\+ 2a)}{\Gamma(N\+ 2a) \over \Gamma(N\+ 1)}
   \right\}^{1/2} (2s)^{-a} \ ,
  \\
   p(s)
  & \underset{s \to \infty}{\sim}
   {1\over \Gamma(a\+\thalf)} \left({2a\+ 1 \over 2a\m 1}\right)^{1/4}
   \left\{ \sqrt{N(N\+ 2a)}{\Gamma(N\+ 2a\+ 1) \over \Gamma(N)}
   \right\}^{1/2} (2s)^{-a-1} \ .
\end{split}
\label{Dcauchy-pq-bc}
\end{equation}
\end{proposition}
Proof - One starts by recasting (\ref{Dcauchy-sde:e}) for $ u $ in terms of
$ q $ and the primary quantities, namely
\begin{equation}
  [\beta_0-u(2a\m 1)] = {q \over sR_0}
  \left[ (1\+ s^2)q'+asq-2(1\+ s^2)qR_0 \right] \ ,
\end{equation}
and similarly for $ [\gamma_0+w(2a\+ 1)] $ by using this result and the 
integral (\ref{Dcauchy-int}). One can then eliminate $ u, w $ using these
expressions in (\ref{Dcauchy-start}) thus relating $ q, q' $ solely in terms
of $ \sigma, R_0 $ by a quadratic equation, and arriving at (\ref{Dcauchy-pq}).
The case for $ p, p' $ is virtually the same. The boundary conditions for 
$ q, p $ follow from the asymptotic forms of $ \phi(s), \psi(s) $ as 
$ s \to \infty $ and their connection with Jacobi polynomials, 
(\ref{phi-psi},\ref{cauchy-poly}). The sign of the radical is determined by a
comparison of the boundary conditions for $ q, p $ and $ \sigma, R_0 $.
$ \square $.

The other auxiliary quantities $ u, w $ can then be found from quadratures also,
given $ q, p $ using (\ref{Dcauchy-sde:c},\ref{Dcauchy-sde:d}).

\subsection{Reduction to \mbox{Painlev\'e} Transcendents}
The second degree second order equations of Proposition 11 can also be related 
to \mbox{Painlev\'e} transcendents.
\begin{proposition}
The variables $ R_{0}(s) $, $ F(s) $, and $ \sigma(s) $ are given in terms of 
the \mbox{Painlev\'e-VI} transcendent $ \omega(x) $ by the formulae, 
\begin{align}
  -2iR_{0} &\,=\, \frac{\ep_1 2x(x - 1)\omega' 
                  - (\omega - 1)\bigl\{ \ep_1(\omega + x) - 2ax \bigr\}}
                    {2\sqrt{x}(x - 1)\omega}\,,  \\
  -iF &\,=\, \frac{\ep_1 2x(1 - x)\omega' + \ep_1 (\omega - 1)(\omega + x) 
             + 2ax}{2\sqrt{x}\omega}\,,  \\
  i\sigma &\,=\, \null - \frac{\bigl\{ 2x(x - 1)\omega' - (\omega - 1)(\omega + x) 
                 \bigr\}^2} {8\sqrt{x}\omega(\omega - 1)(\omega - x)}  \nonumber \\*
          &\qquad\qquad\null \,+\, \frac{\sqrt{x}(\omega - 1)\bigl\{ N(N\+ 2a)\omega 
                 + a^{\,2}x \bigr\}} {2\omega(\omega - x)}\,,
\end{align}
where \mbox{$\ep_1 := \pm 1$} and $ s = i\sqrt{x} $, and with the parameters,
\begin{equation}
\al = \tfrac{1}{8}\,, \qquad 
     \be = \null - \tfrac{1}{8}(1 \mp 2a)^2, \qquad 
     \ga = 0, \qquad 
     \de = \tfrac{1}{2}(1 - [N\+ a]^2).
\end{equation}
\end{proposition}
Proof - Under the transformation $ s \mapsto -is $ and 
$ R_{0} \mapsto -2iR_{0} $
the equation (\ref{Dcauchy-ode2b}) is identical in form to equation (5.14) in 
\cite{WFC-00}, and consequently the rest of the analysis there carries
through with an adjustment of the transcendent parameters.
$ \square $

\subsection{Special Cases for low $ N $}
In this part we present the analogous results for the first two finite-$ N $ 
cases $ N = 1 $ and $ N = 2 $ by direct calculation of the probability 
$ E_2(0;I) $.
\begin{proposition}
The probability (\ref{CyUE-gap:b}) for $ N = 1 $ and the associated functions 
are given by
\begin{equation}
 \begin{split}
  E_2(0;I)
  & =  {2\Gamma(a\+ 1) \over \sqrt{\pi}\Gamma(a\+\half)}
       s\, {}_{2}F_{1}(a\+ 1,\half;\thalf;-s^2)
  \ , \\
  & =  1 - {\Gamma(a\+ 1) \over \sqrt{\pi}\Gamma(a\+\thalf)}
       s^{-2a-1}\, {}_{2}F_{1}(a\+ 1,a\+\half;a\+\thalf;-s^{-2})
  \ , \\
  \sigma(s)
  & = {(1\+ s^2)^{-a} \over 2s\, {}_{2}F_{1}(a\+ 1,\half;\thalf;-s^2)}
  \ , \\
  F(s)
  & = as
      + {(1\+ s^2)^{-a} \over s\, {}_{2}F_{1}(a\+ 1,\half;\thalf;-s^2)}
  \ ,
\end{split}
\label{Dcauchy-N1}
\end{equation}
and for the corresponding probability with $ N = 2 $ is
\begin{equation}
 \begin{split}
  E_2(0;I)
  & =  {4\Gamma(a\+ 2)\Gamma(a\+ 1) \over \pi\Gamma(a\+\half)\Gamma(a\+\thalf)}
       s^2 {}_{2}F_{1}(a\+ 2,\half;\thalf;-s^2)
       \left[ {}_{2}F_{1}(a\+ 1,\half;\thalf;-s^2)-(1\+ s^2)^{-a-1} \right]
  \ , \\
  & =  \left\{ 1 - 
               {\Gamma(a\+ 2) \over \sqrt{\pi}\Gamma(a\+\fhalf)}
               s^{-2a-3}\, {}_{2}F_{1}(a\+ 2,a\+\thalf;a\+\fhalf;-s^{-2})
       \right\}
  \\
  & \phantom{=} \times \left\{ 1 - 
               {2\Gamma(a\+ 2) \over \sqrt{\pi}\Gamma(a\+\thalf)}
               s^{-2a-1}\, {}_{2}F_{1}(a\+ 1,a\+\half;a\+\thalf;-s^{-2})
                       \right\}
  \ , \\
  \sigma(s)
  & = {(1\+ s^2)^{-a-1} \over 2s} \left\{
      {2(a\+ 1)s^2 \over {}_{2}F_{1}(a\+ 1,\half;\thalf;-s^2)-(1\+ s^2)^{-a-1}}
    + {1 \over {}_{2}F_{1}(a\+ 2,\half;\thalf;-s^2)}
                                 \right\}
  \ , \\
  F(s)
  & = as + {(1\+ s^2)^{-a-1} \over s}  \left\{
      {2(a\+ 1)s^2 \over {}_{2}F_{1}(a\+ 1,\half;\thalf;-s^2)-(1\+ s^2)^{-a-1}}
    - {1 \over {}_{2}F_{1}(a\+ 2,\half;\thalf;-s^2)}
                                 \right\}
  \ ,
\end{split}
\label{Dcauchy-N2}
\end{equation}
in terms of the Gauss hypergeometric function $ {}_{2}F_{1}(a,b;c;z) $.
\end{proposition}
Proof - We proceed in a parallel manner as in the proof of Proposition 6
and the relations given there.
$ \square $

Again one can show that these two specific cases are solutions of the second 
order differential equations (\ref{Dcauchy-ode2a},\ref{Dcauchy-ode2b}) and 
satisfy the boundary conditions (\ref{Dcauchy-bca},\ref{Dcauchy-bcb}), 
using the contiguous and differentiation formulae for the Gauss hypergeometric 
function.
 
\subsection{The Thermodynamic Limit for the Circular Jacobi Ensemble}
In Section 4.3 we saw that the stereographic projection (\ref{stereo}) maps,
in the limit $ L \to \infty $, the interval $ (s,\infty) $ in the Cauchy
ensemble to the interval $ (0,x) $ in the scaled circular Jacobi ensemble.
Similarly this mapping and limiting procedure maps the region
$ (\infty,-s) \cup (s,\infty) $ in the Cauchy ensemble to the interval
$ (-x,x) $ in the scaled circular Jacobi ensemble. Thus
\begin{equation}
   E_{2}(0;(-\infty,-s) \cup (s,\infty);(1\+ t^2)^{-N-a};N) =
   E_{2}(0;(-x,x);|1\m z|^{2a};N) \ ,
\label{Dcauchy-prob-xfm}
\end{equation}
and since with $ \sigma(t) $ specified by Proposition 11
\begin{equation}
   E_{2}(0;(-\infty,-s) \cup (s,\infty);(1\+ t^2)^{-N-a};N) =
   \exp\left(-2\int^{\infty}_{s} dt\;{\sigma(t) \over 1\+ t^2}\right) \ ,
\label{Dcauchy-prob-int}
\end{equation}
we have
\begin{equation}
   E_{2}(0;(-x,x);|1\m e^{2\pi iy/L}|^{2a};N) =
   \exp\left(-{2\pi \over L}\int^{x}_{0} dy\;\sigma(\cot\pi y/L)\right) \ .
\label{Dcirc-prob-int}
\end{equation}
Furthermore, with
\begin{align}
   E_{2}(0;(-x,x);a) 
  & = \lim_{\overset{N,L \to \infty}{N/L = \rho}}
       E_2(0;(-x,x);|1\m e^{2\pi iy/L}|^{2a};N) \ ,
  \label{Dprob-limit:a} \\
   {1 \over N}\sigma(\cot\pi y/L)
  & \mapsto \sigma(\pi\rho y)
  \label{Dprob-limit:b}
\end{align}
(\ref{Dcirc-prob-int}) gives
\begin{equation}
   E_{2}(0;(-x,x);a) =
   \exp\left(-2\int^{\pi\rho x}_{0} dy\;\sigma(y)\right) \ .
\label{Dscale-prob-int}
\end{equation}

It is of relevance to consider the scaled form of the Cauchy kernel under
the stereographic projection (\ref{stereo}). Using the relation between the
Jacobi polynomial and the hypergeometric function $ {}_2F_1 $, transformations
and confluent forms of the latter, and the relationship of $ {}_1F_1 $ to the
Bessel function one can show \cite{F-00}
\begin{equation}
  K_N(s_1,s_2) ds_2 \to K(x_1,x_2) dx_2 \ ,
\label{scale-kernel}
\end{equation}
where
\begin{equation}
   K(x,y) = (\pi\rho x)^{1/2}(\pi\rho x)^{1/2}
   { J_{a+1/2}(\pi\rho x)J_{a-1/2}(\pi\rho y)
     - J_{a+1/2}(\pi\rho y)J_{a-1/2}(\pi\rho x) \over 2(x-y) } \ .
\label{Bessel-kernel}
\end{equation}
This particular Bessel kernel was first identified in \cite{NS-93}. It has
been used in \cite{FO-96} to provide a direct derivation of 
(\ref{Dscale-prob-int}), and furthermore specify $ \sigma(y) $ therein as 
the solution of a particular second order ODE. This latter result can be
considered as a limiting case of (\ref{Dcauchy-ode2a}).
We summarise here the known results for this kernel, for use later on, and
whilst we use the same symbols for primary and auxiliary quantities in this
case and for the finite Cauchy problem the reader should not confuse them.
We intend to display the interconnections between the scaled Cauchy 
quantities and the corresponding Bessel kernel ones in the following subsection.

\begin{proposition}
The gap probability and associated quantities for the Bessel kernel on the
interval $ (-x,x) $ with
\begin{equation}
\begin{split}
  \phi(x) & = \left(x/2\right)^{1/2}J_{a+1/2}(x) \ ,
          \nonumber\\
  \psi(x) & = \left(x/2\right)^{1/2}J_{a-1/2}(x) \ ,
          \nonumber
\end{split}
\label{Bessel-phi-psi}
\end{equation}
are defined by the following differential equations with 
$ R \equiv R(x,x) $
\begin{align}
  u'
  & = 2q^2 \ ,
  \label{Bessel-sde:c}\\
  w'
  & = 2p^2 \ ,
  \label{Bessel-sde:d}\\
  xq'
  & =   xp
      + (-a+[u-w]) q \ ,
  \label{Bessel-sde:e}\\
  xp'
  & = - xq
      + ( a-[u-w]) p \ ,
  \label{Bessel-sde:f}\\
  xR
  & = x(q^2+p^2)
     + 2(-a+[u-w]) qp
     + 2(qp)^2  \ ,
  \label{Bessel-sde:g}\\
  (xR)'
  & = q^2+p^2 \ .
  \label{Bessel-sde:h}
\end{align}
along with the boundary conditions 
\begin{equation}
\begin{split}
  q(x) & \underset{x \to 0}{\sim} 
         {1\over \Gamma(a\+\thalf)}(\half x)^{a+1} \ ,
       \\
  p(x) & \underset{x \to 0}{\sim} 
         {1\over \Gamma(a\+\half)}(\half x)^{a} \ .
\end{split}
\label{Bessel-bc}
\end{equation}
\end{proposition}
Proof - See reference \cite{FO-96}, with the mapping 
$ x \mapsto \pi\rho x $.
$ \square $

For future reference we note the following integral arising from 
(\ref{Bessel-sde:c},\ref{Bessel-sde:d},\ref{Bessel-sde:e},\ref{Bessel-sde:f}),
\begin{equation}
   2qp = w-u \ .
\label{Bessel-int}
\end{equation}

\begin{proposition}
Consider the scaled function $ \sigma(y) $ in (\ref{Dprob-limit:b}), and
write $ \sigma_1(r) \equiv -2x\sigma(x) $, $ r \equiv 2x $. Then we have that
$ \sigma_1 $ satisfies the ODE
\begin{equation}
   (r\sigma_1'')^2 + 4[-a^2-\sigma_1+r\sigma_1']
    \left\{ (\sigma_1')^2 
            - \left[ a-\sqrt{a^2+\sigma_1-r\sigma_1'} \right]^2
    \right\} = 0 \ ,
\label{Dscale-ode2}
\end{equation}
subject to the boundary conditions
\begin{equation}
   \sigma_1(r) \underset{r \to 0}{\sim}
   -{2 \over \Gamma(a\+\half)\Gamma(a\+\thalf)}
    (\tfrac{1}{4}r)^{2a+1}\ .
\label{Dscale-bc}
\end{equation}
\end{proposition}
Proof - The result comes from the leading order terms in the ODE
(\ref{Dcauchy-ode2a}) after the change of variables (\ref{stereo}) and 
introducing the dependent function (\ref{Dprob-limit:b}).
$ \square $

This is precisely the equation for $ \sigma_1(2x) \equiv -2xR(x,x) $ that 
was derived from the coupled differential equations 
(\ref{Bessel-sde:a}-\ref{Bessel-sde:h}) in \cite{FO-96}. We now want to 
indicate the precise relationships between the scaled Cauchy ensemble quantities
$ \sigma, R_0, q, p $ and the Bessel kernel quantities, which we
shall distinguish by affixing the subscript $ \infty $, namely the quantities
$ R_{\infty}, q_{\infty}, p_{\infty} $. From the arguments in the proposition 
above we see that
\begin{equation}
   \sigma(x) = R_{\infty}(x) \ ,
\label{Dscale-map1}
\end{equation}
after recognising that
\begin{equation}
   E_2(0;(-x,x);a) = \exp\left( -2\int^{\pi\rho x}_0 dy\,
           R_{\infty}(y) \right) \ .
\label{Bessel-sde:a}
\end{equation}
Having made this identification we can combine (\ref{Bessel-sde:g},
\ref{Bessel-sde:h}) according to
\begin{equation}
    x{d \over dx}\sigma_1 - \sigma_1 = - 4a(q_{\infty}p_{\infty})
                                       - 4(q_{\infty}p_{\infty})^2 \ .
\label{Bessel-map2}
\end{equation}
Then by performing the scaling limit on (\ref{Dcauchy-sde:h}) and comparing
with the above equation we find a simple scaling for $ R_0 $
\begin{equation}
   NR_0(s) = r(\pi\rho x) \ ,
\label{Bessel-map3}
\end{equation}
along with the identification in the thermodynamic limit
\begin{equation}
   r(x) = xq(x)p(x) = xq_{\infty}(x)p_{\infty}(x) \ .
\label{Bessel-map4}
\end{equation}
Here we define the scaled auxiliary variables by
$ q(s) \mapsto q(\pi\rho x) $ and $ p(s) \mapsto p(\pi\rho x) $.
Utilising these two simple scalings we have the following result.
\begin{proposition}
The scaled limits of the auxiliary quantities for the Cauchy ensemble on the
double interval $ q(x), p(x) $ are related to the auxiliary variables for
the Bessel kernel $ q_{\infty}(x), p_{\infty}(x) $ by 
\begin{align}
   q(x) 
  & = {1\over \Gamma(a\+\half)}\left({2a\m 1 \over 2a\+ 1}\right)^{1/4}
      \left(\half x\right)^{a}
      \exp\left( -\int^x_0 dy\; {q_{\infty}(y) \over p_{\infty}(y)}
          \right) \ ,
  \label{Dcauchy-map:q}\\
   p(x) 
  & = {1\over \Gamma(a\+\thalf)}\left({2a\+ 1 \over 2a\m 1}\right)^{1/4}
      {\left(\half \right)^{a+1} \over x^a}
      \exp\left( +\int^x_0 dy\; {p_{\infty}(y) \over q_{\infty}(y)}
          \right) \ .
  \label{Dcauchy-map:p}
\end{align}
\end{proposition}
Proof - Utilising the scaling forms (\ref{Dprob-limit:b},\ref{Bessel-map3}) 
in the expressions for the logarithmic derivatives of $ q $ in
(\ref{Dcauchy-pq}) we find
\begin{equation}
   {q' \over q} - ax^{-1} = 
   -{q_{\infty} \over p_{\infty}} \quad\text{or}\quad
   -{p_{\infty} \over q_{\infty}} \ . 
\end{equation}
The correct choice can be made and the integration constant found by imposing 
both the boundary conditions (\ref{Dcauchy-pq-bc},\ref{Bessel-bc}). The
situation for $ p $ is entirely analogous.
$ \square $

\subsection{Reduction of scaled Limit to \mbox{Painlev\'e} Transcendents}

In this part we show how to reduce the auxiliary variables $ q, p $ (we
now drop the subscript $ \infty $) for the Bessel kernel directly to 
\mbox{Painlev\'e} transcendents. 
\begin{proposition}
The variable $ p(x) $ for the Bessel function kernel with parameter $ a $ on
the interval $ (-x,x) $ is related to the \mbox{Painlev\'e-V} transcendent 
$ y(x) $ by
\begin{equation}
   p = -\sqrt{x \over 2}{1+y \over 1-y} \ ,
\end{equation}
whose parameter values are
\begin{equation}
   \alpha  =  \tfrac{1}{32}(1-2a)^2 \ ,\quad
   \beta   = -\tfrac{1}{32}(1-2a)^2 \ ,\quad
   \gamma  =  0 \ ,\quad
   \delta  = -2  \ .
\end{equation}
In terms of this transcendent $ q(x), \sigma_1(2x) $ are
\begin{align}
   q & = \sqrt{2 \over x}{1 \over 4y}
            \left[-x\dot{y} + \tfrac{1}{4}(2a\m 1)(1-y^2) \right] \ , \\
   \sigma_1 & =
   {1\over y}\left[ {x\dot{y} \over 1\m y} + \quar(1+y) \right]^2
  -\quar a^2{(1+y)^2\over y} - x^2\left({1+y \over 1-y}\right)^2 \ .
\end{align}
\end{proposition}
Proof - We eliminate all variables other than $ q $ and $ p $ from equations
(13)-(17) in \cite{FO-96} using (\ref{Bessel-int}), and employ this with the 
coupled set (\ref{Bessel-sde:e},\ref{Bessel-sde:f}). Elimination of $ q $ using
\begin{equation}
  q = {xp' - ap \over 2p^2 - x} \ ,
\end{equation}
leads to the ODE for $ p $,
\begin{equation}
  p'' = {2p \over 2p^2 - x}(p')^2 - {2p^2 \over x(2p^2-x)}p'
        + {p \over x}(2p^2 - x)
        + a(1\m a){p \over x(2p^2 - x)} \ .
\end{equation}
Then applying the above transformation reduces this to a \mbox{Painlev\'e-V}.
$ \square $

\vfill\eject

\section{Conclusions}

The circular Jacobi ensemble describes a degeneracy of order $ a $ in the
spectrum of unitary random matrices. We have found the probability that the
one-sided interval $ (0,x) $ to the right of the singularity is eigenvalue
free, and similarly for the symmetrical interval $ (-x,x) $ about the 
singularity. The former probability has been expressed in terms of a
\mbox{Painlev\'e-VI} transcendent, which in the scaled limit reduces to a
\mbox{Painlev\'e-V} transcendent. These calculations were performed by first
mapping the circular Jacobi ensemble, via a stereographic projection, to the
Cauchy ensemble of random Hermitian matrices. This latter ensemble is 
classical, having the same essential properties as the more familiar Hermite,
Laguerre and Jacobi classical ensembles. As the probability of a gap free 
interval about an endpoint of these latter three ensembles are known in terms
of Painlev\'e type transcendents from earlier studies \cite{TW-94,HS-99},
as is the probability for two intervals at the ends of the support in the
Hermite and symmetric Jacobi cases \cite{WFC-00}, by studying the Cauchy 
ensemble we are completing a classification program of determining the
Painlev\'e transcendents associated with such probabilities in the classical
ensembles.

The Cauchy ensemble is closely related to the symmetric Jacobi ensemble. 
Previous studies of the probabilities of gap free regions in this latter
ensemble \cite{TW-94,HS-99,WFC-00} have encountered third order differential
equations, which through various means have been integrated to second order
equations. In contrast in this study we are able to find an extra integral
of the coupled differential equations, unknown from previous studies, which 
provides a direct path to the second order differential equation 
characterising the gap probability. We consider that the method of constructing 
this new integral will apply more generally than just to the ensemble treated 
here and will have important implications for the application of the Tracy
and Widom formalism to other problems.

The case $ a = 1 $ of the scaled circular Jacobi ensemble has particular
relevance to the scaled CUE of random unitary matrices, or equivalently to
the scaled GUE of random Hermitian matrices. Thus it corresponds to fixing
an eigenvalue at the origin in these latter two scaled ensembles. The
probability of no eigenvalues in the one-sided interval $ (0,x) $ of the
scaled circular Jacobi ensemble is then proportional to the derivative of
the probability that there are no eigenvalues in the interval
$ (0,x) $ of the scaled CUE or GUE. This is turn allows us to derive the
exact Wigner surmise type form (\ref{p2-wigner}) for the spacing
probability between consecutive eigenvalues in the bulk of matrix ensembles
with unitary symmetry.

\bibliographystyle{siam}
\bibliography{moment,random_matrices,nonlinear}
\vspace*{30mm}
\begin{center}
\bf Acknowledgements
\end{center}
Both authors acknowledge the support of an Australian Research Council 
Grant whilst this work was performed,
and NSW thanks Andrew Hone and Christopher Cosgrove for discussions.
\vfill\eject

\end{document}